\documentclass[pra,aps,twocolumn,superscriptaddress,showpacs,showkeys]{revtex4-1}
\usepackage{amsmath,amsthm,amssymb,graphicx,subfigure,xcolor}
\usepackage[unicode=true]{hyperref}
\hypersetup{
     colorlinks=true,       		
     linkcolor=blue,          	
     citecolor=red,            
     urlcolor=magenta,           	
 }

\newtheorem*{theorem*}{Theorem}

\newtheorem*{corollary*}{Corollary}

\newtheorem*{lemma*}{Lemma}

\newtheorem*{proposition*}{Proposition}
\theoremstyle{definition}

\newtheorem*{definition*}{Definition}
\theoremstyle{remark}

\newtheorem*{remark*}{Remark}

\newcommand{\ket}[1]{|#1\rangle}

\begin{document}

\title{Quantum backflow in solutions to the Dirac equation of the spin-$1/2$ free particle}

\author{Hong-Yi Su}
\email{hysu@gscaep.ac.cn}
\affiliation{Graduate School of China Academy of Engineering Physics, Beijing 100193, People's Republic of China}
\affiliation{Department of Physics Education, Chonnam National University, Gwangju 500-757, Republic of Korea}

\author{Jing-Ling Chen}
\email{chenjl@nankai.edu.cn}
\affiliation{Theoretical Physics Division, Chern Institute of Mathematics, Nankai University, Tianjin 300071, People's Republic of China}
\affiliation{Centre for Quantum Technologies, National University of Singapore, 3 Science Drive 2, Singapore 117543}

\date{\today}
\begin{abstract}
It was known that a free, nonrelativistic particle in a superposition of positive momenta can, in certain cases, bear a negative probability current --- hence termed quantum backflow. Here, it is shown that more variations can be brought about for a free Dirac particle, particularly when negative-energy solutions are taken into account. Since any Dirac particle can be understood as an antiparticle that acts oppositely (and vice versa), quantum backflow is found to arise in the superposition (i) of a well-defined momentum but different signs of energies, or more remarkably (ii) of different signs of both momenta and energies. Neither of these cases has counterpart in nonrelativistic quantum mechanics. A generalization by using the  field-theoretic formalism is also presented and discussed.
\end{abstract}

\pacs{03.65.Ta, 03.65.Pm}

\keywords{Quantum backflow; relativistic wavefunction; spin-1/2 particle.}

\maketitle

\section{Introduction}
There has been known to exist various ways to measure the \emph{velocity} of a classical wave, such as the phase velocity ${\rm v}_p=\omega/k$, the group velocity ${\rm v}_g={\rm d}\omega/{\rm d}k$, etc., with $\omega$ and $k$ respectively denoting the angular frequency and the wavenumber. In a different context, for a matter wave described by Schr\"odinger's or Dirac's equation, an effective velocity, defined as
\begin{equation}
  {\rm v}=\frac{\nabla S}{m}=\frac{j}{\rho},\label{bohm-velo}
\end{equation}
where $S$ denotes the \emph{quantum} Hamilton-Jacobi function~\cite{bohm-undivided}, and density $\rho$ and current $j$ are governed by the probability conservation equation, is often to serve as an important factor to evaluate many dynamical properties of a quantum system~\cite{landau}. It is notable that with the quantum dispersion relations the ${\rm v}_p$ and the ${\rm v}_g$ for free particles should, if nonvanishing, have the same sign~\cite{samesign}.

Nevertheless this is not true of Eq.~(\ref{bohm-velo}) with all free-particle states, particularly with coherent ones that have no well-defined momentum. For instance, given the simplest superposed state
\begin{equation}
  \ket{\psi}=\ket{k_1}+a\ket{k_2},\label{superposedstate}
\end{equation}
with $k_1=0,~k_2=k>0$ for simplicity, it was found that a peculiar phenomenon, termed \emph{quantum backflow}, could happen. That is, Eq.~(\ref{bohm-velo}), which is equal, in the position representation, to
\begin{equation}
  {\rm v}=\frac{\hbar}{m}{\rm Im}[\frac{\partial_x\psi}{\psi}]=\frac{a\hbar k}{m}\frac{a+\cos kx}{1+a^2+2a\cos kx},\label{berrysresult}
\end{equation}
may take negative values with certain $a$ and $k$~\cite{Berry}, while the expectation of momentum $\langle\hat p\rangle$ remains positive.

\begin{figure}[t]
\includegraphics[width=70mm]{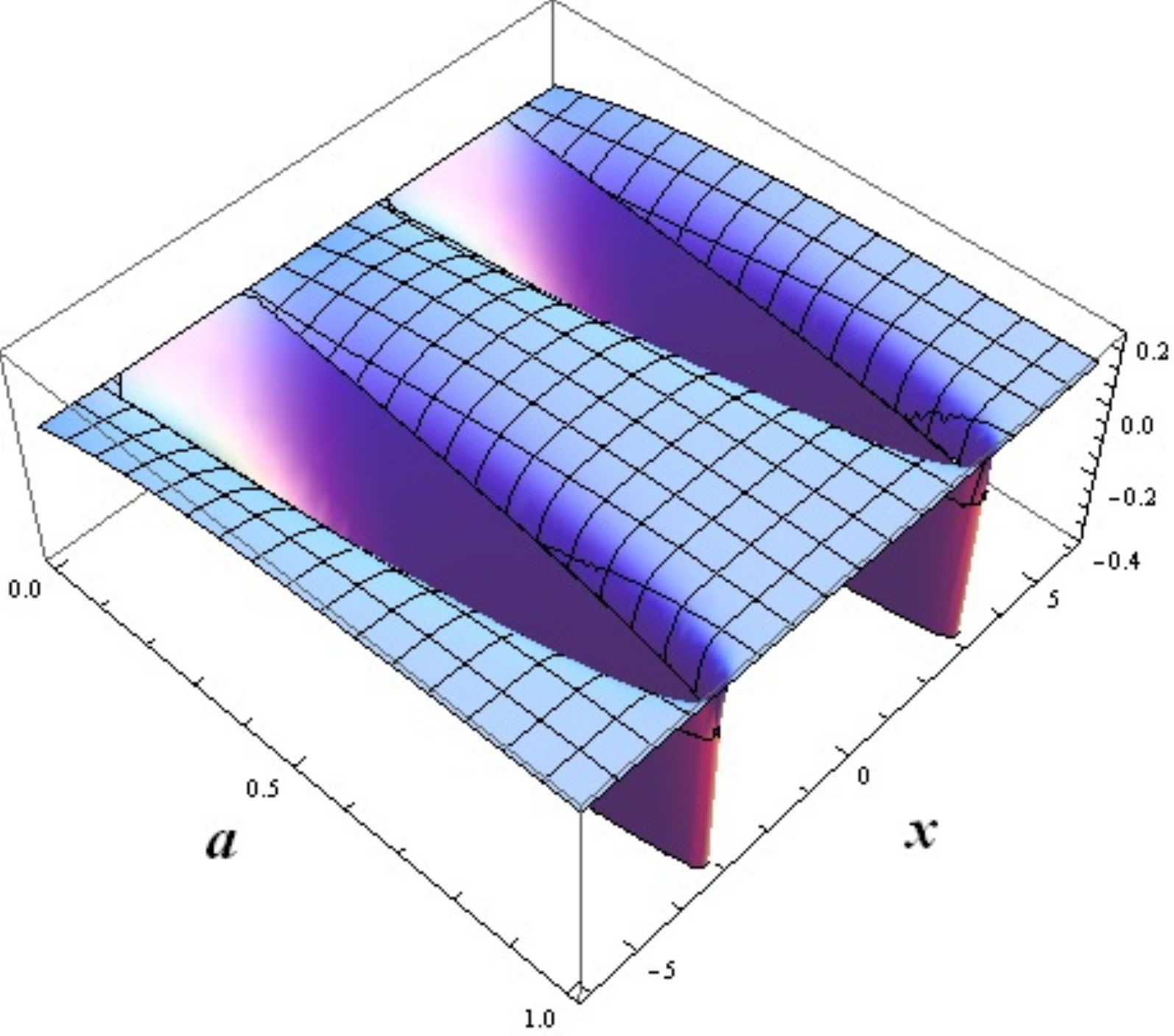}\\
\caption{Quantum backflow and quantum potential. For simplicity, we choose $m=\hbar=k=1$, within $a\in[0,1]$ and $x\in[-6,6]$. The unmeshed and the meshed regions indicate ${\rm v}<0$ and $Q>0$ derived in (\ref{berrysresult}) and (\ref{qpotential}), respectively.
  }\label{fig1}
\end{figure}

This phenomenon can be well understood in, e.g., the pilot-wave theory~\cite{madelung27,debroglie27,bohm52} --- an ontological interpretation of quantum mechanics. Accordingly, the ${\rm v}$ in (\ref{bohm-velo}) should be understood as the velocity of a particle that is moving along a certain trajectory. In this case, the particle in state (\ref{superposedstate}), while not subject to any classical potential, is indeed subject to a nonzero \emph{quantum potential}~\cite{bohm-undivided}
\begin{equation}
  Q=-\frac{\hbar^2}{2m}\frac{\nabla^2R}{R}=\frac{a\hbar^2 k^2}{2m}\frac{(1+a\cos kx)(a+\cos k x)}{(1+a^2+2a\cos kx)^2},\label{qpotential}
\end{equation}
with $R=\sqrt{\psi\psi^*}$. This potential, when positive, may reflect the particle backward, therefore leading to the backflows (see also Fig.~\ref{fig1}). (It is also related to the arrival-time problems~\cite{allcock69,muga2000}.) The standard momentum $\langle\hat p\rangle$ remains unchanged since it does not take account of such potentials.

Let us stress that the phenomenon, that a superposition of components with some positive properties yield an opposite collective movement, is not unique in quantum theory; similar phenomena --- e.g., the phase velocity versus the group velocity --- do exist in the field of classical optics. But there are distinctions: (i) in classical optics, $\omega$ and $k$ in each component are independent quantities, but in quantum mechanics they are connected by the dispersion relations~\cite{samemass}; (ii) the occurrence of backflows not only depends on the position --- it could be that it may happen in some region but may not elsewhere --- but is also time changing (see, e.g.,~\cite{Berry}); (iii) most importantly, quantum mechanical wavefunctions, according to Born's rule, have a particle-wave duality, while classical waves do not.

By far, there have been many results on the backflow~\cite{Berry,bracken1994probability} and its related phenomena~\cite{allcock69,muga2000}, such as the weak value~\cite{Berry} and the superoscillation~\cite{berry94,berry2009}. Yet, less is known about its relativistic extensions, except for just a few~\cite{melloy1998the,melloy1998probability,berry2012}. Particularly, in~\cite{berry2012}, it was shown that for Klein-Gordon and Dirac waves, which consist of superpositions of many plane waves, the local group velocity can exceed the speed of light $c$, yielding potentially observable phenomena outside the spectrum of the plane wave. This type of superluminality involves very similar concepts as quantum backflow. Results in the present paper share similarities with those in~\cite{berry2012}:
\begin{itemize}
  \item  Objects discussed are massive quantum particles.

  \item All constituent plane waves are positive and subluminal.

  \item One-dimensional waves are considered.
\end{itemize}
On the other hand, there are a few differences:
\begin{itemize}
  \item Rather than weak values, we are focused on quantum backflow, which is also related to the Zitterbewegung.

  \item We use four-component spinors with more degrees of freedom.

  \item We consider only two waves in a superposition.

  \item Dynamical behaviors were also investigated in~\cite{berry2012}.
\end{itemize}
Thus, the present paper can fairly be regarded as a timely extension of~\cite{berry2012}.

It is very meaningful to investigate the relativistic effects upon quantum backflow. For one thing, relativistic quantum mechanics provides more degrees of freedom in the system than that with only orbital part of wavefunctions considered. For another, concepts like negative energies, spins, helicities and negative-positive energy transition~\cite{dirac28,superpose} have been widely believed to be responsible for a few unique quantum phenomena, such as the beam split in the Stern-Gerlach experiment, the spin-orbital angular momentum conservation, the Zitterbewegung, etc. Thus, more studies on how relativistic features affect quantum backflow are inevitable.

It is notable that in measuring, for instance, the momentum $k_i$, one needs to have the state $\ket{\psi}=\sum_i a_i\ket{k_i}$ coupled with a meter initially in $\ket{0}$, then, according to time evolution, the combined system becomes entangled, i.e., $\ket{\psi}\otimes\ket{0}\rightarrow\sum_i a_i\ket{k_i}\otimes\ket{i}$, such that the $i$th state of the meter, $\ket{i}$, with a probability $|a_i|^2$, indicates $\ket{k_i}$ of the system. Hence, for different components $\ket{k_i}\otimes\ket{i}$'s, one must require the combined momenta to be equal to one another, though $\ket{k_i}$'s alone are clearly not. The same argument for concerns of conservation applies to other quantities like energy, angular momentum, etc.

\section{Quantum backflow of the free Dirac particle for spin $\frac{1}{2}$}

Since ${\rm v}$ defined in (\ref{bohm-velo}) is proportional to $j$ just up to a positive quantity $\rho$, in studying backflows below, we are focused on $j$ instead of ${\rm v}$, and on its direction fluxing forward or backward. The Dirac current $j$ is then written as
\begin{equation}
   j=c\Psi^\dagger\alpha\Psi.\label{rel-current}
\end{equation}
Here for Dirac's Hamiltonian $H=c\alpha\cdot p+\beta m c^2$, we take the Pauli-Dirac representation:
\begin{equation}
  \alpha=\left[\begin{matrix}
     &  \sigma\\ \sigma &
  \end{matrix}\right],~~ \beta=\left[\begin{matrix}
    \textbf{1} &   \\  & -\textbf{1}
  \end{matrix}\right],~~\Sigma=\left[\begin{matrix}
    \sigma &   \\  & \sigma
  \end{matrix}\right],
\end{equation}
where $\textbf{1}$ denotes the identity matrix, $\Sigma$ defines the spin, and $\sigma=(\sigma_x,\sigma_y,\sigma_z)$ with
\begin{equation}
\begin{split}
  \sigma_x=\left[\begin{matrix}
     &  1\\ 1 &
  \end{matrix}\right],~~\sigma_y=\left[\begin{matrix}
     &  -i\\ i &
  \end{matrix}\right],~~\sigma_z=\left[\begin{matrix}
    1 &  \\  & -1
  \end{matrix}\right].
\end{split}
\end{equation}

There is a natural connection between the relativistic backflow and the Zitterbewegung. It is noted that the absolute value of Dirac's velocity
\begin{equation}
  \frac{{\rm d}r}{{\rm d}t}=-\frac{i}{\hbar}[r,H]=c~\alpha
\end{equation}
is independent of the orbital velocity of the particle and is always equal to the speed of light. This is related to the Zitterbewegung. An orbital-dependent velocity, however, may be obtained by an appropriate projection to the positive energy space.

To proceed, we choose observables $H,~p$ and $\Sigma\cdot p$, commuting with one another, to solve
$H\Psi=E\Psi$. By further defining
\begin{equation}
  \Psi=\frac{1}{\sqrt{A}}\left[ \begin{matrix}
    \varphi \\ \chi
  \end{matrix} \right],~~\chi=\frac{c\sigma\cdot p}{E+mc^2} \varphi,~~A=\Psi^\dagger\Psi,
\end{equation}
and taking $p=(0,0,\hbar k)$ for simplicity, we then get $\sigma\cdot p=\lambda\hbar k$ with $\lambda=\pm1$, so that for a certain $k$ the solutions read
\begin{align}
\Psi(E_+,k,\lambda) &= \biggr(\frac{2|E|}{|E|+mc^2}\biggr)^{-1/2}\left[\begin{matrix} \varphi_\lambda \\ \frac{\lambda c\hbar k}{|E|+mc^2}\varphi_\lambda \end{matrix}\right]e^{ik z},\\
\Psi(E_-,k,\lambda) &= \biggr(\frac{2|E|}{|E|+mc^2}\biggr)^{-1/2}\left[\begin{matrix} \frac{-\lambda c\hbar k}{|E|+mc^2}\varphi_\lambda \\ \varphi_\lambda \end{matrix}\right]e^{ik z},
\end{align}
where $\varphi_{+1}=[ 1, 0]^{\rm T}$, $\varphi_{-1}=[ 0, 1]^{\rm T}$, and $E=E_{\pm}=\pm\sqrt{c^2 \hbar^2 k^2+m^2 c^4}$.

Given $k>0$, we now list a summary of our results with a variety of free-particle states~\cite{unorthogonal}:
\begin{enumerate}

\item\label{eigen} $\Psi=\Psi(E_{\pm},k,\lambda)$:
\begin{equation}
  \begin{split}
  j_x=j_y=0,~~j_z=\frac{c^2\hbar k}{|E|}\times{\rm Sgn}(E_{\pm}),
  \end{split}
\end{equation}
where ${\rm Sgn}(E)$ denotes the sign function of support $E$.

\item\label{nonzero00} $\Psi=[\Psi(E_+,k,\lambda)+a e^{i \phi}\Psi(E_-,k,\lambda)]/\sqrt{1+a^2}$:
\begin{equation}
\begin{split}
  j_x&=j_y=0,\\
  j_z&=\frac{c^2}{(1+a^2)|E|}\biggr[\hbar k(1-a^2)+2\lambda a m c \cos\phi \biggr]\\
     &=\frac{m c^2}{(1+a^2)|E|}\biggr[ \gamma v (1-a^2)+2\lambda a c \cos\phi \biggr],
\end{split}
\end{equation}
where $\gamma=1/\sqrt{1-(v/c)^2}$, $v\in[0,1]$.
For instance, we obtain at $\phi=\pi,~\lambda=1$ a negative $j_z$ in the domain
\begin{align}
  a\in(a_c,1], ~~  a_c=\sqrt{\frac{\gamma-1}{\gamma +1}}.
\end{align}

\item\label{posi-nega} $\Psi=[\Psi(E_+,k,\lambda)+a e^{i \phi}\Psi(E_-,-k,\lambda)]/\sqrt{1+a^2}$:
\begin{equation}
\begin{split}
  j_x&=j_y=0,\\
  j_z&=\frac{c^2\hbar k}{|E|}+\frac{2\lambda ac}{1+a^2} \cos(2k z-\phi) \\
  &=v+\frac{2\lambda ac}{1+a^2} \cos(2k z-\phi).
\end{split}
\end{equation}
Similarly to case \ref{nonzero00}, at $\phi=\pi,~\lambda=1$, we get a negative $j_z$ around $z=n\pi/k,~n=0,\pm1,\pm2,\cdots,$ by taking, for instance, $v=ac,~m=1$.

\item\label{nonzero-jx} $\Psi=[\Psi(E_+,k,\lambda)+a e^{i \phi}\Psi(E_-,k,-\lambda)]/\sqrt{1+a^2}$:
\begin{equation}
  \begin{split}
  j_x&=\frac{2ac}{1+a^2}\cos\phi,~~  j_y=\frac{2\lambda ac}{1+a^2}\sin\phi,\\
  j_z&=\frac{c^2\hbar k}{|E|}\frac{1-a^2}{1+a^2}>0.
  \end{split}
\end{equation}
Most intriguingly, here $j_x$ and $j_y$ can be nonvanishing despite that $p_x=p_y=0$.

\item\label{nonzero-jx2} $\Psi=[\Psi(E_+,k,\lambda)+a e^{i \phi}\Psi(E_-,-k,-\lambda)]/\sqrt{1+a^2}$:
\begin{equation}
  \begin{split}
  j_x&=\frac{2a m c^3}{(1+a^2)|E|}\cos(2k z-\phi),\\
  j_y&=\frac{-2\lambda a m c^3}{(1+a^2)|E|}\sin(2k z-\phi),\\
  j_z&=\frac{c^2\hbar k}{|E|}>0.
  \end{split}
\end{equation}
Similarly to case \ref{nonzero-jx}, $j_x$ and $j_y$ can be nonvanishing despite that $p_x=p_y=0$.

\item\label{zero_nonzero1} $\Psi=\Psi(E_\pm,0,\lambda)+a\{2|E|/(|E|+m)\}^{1/2}\Psi(E_\pm,k,\lambda)$:
\begin{equation}
  \begin{split}
  j_x&=j_y=0,\\
  j_z&=\pm \frac{ac^2\hbar k}{|E|+mc^2}\biggr[ a+\cos(k z+\phi) \biggr].
  \end{split}
\end{equation}
$j_z$ can be negative around $z=2n\pi/k,~n=0,\pm1,\pm2,\cdots,$ for both the positive-energy case, for instance at $\phi=\pi,~\lambda=1,~a=1/\sqrt{2},~m=1$, and the negative-energy case, for instance at $\phi=0,~\lambda=1,~a=1/\sqrt{2},~m=1$.

\item\label{zero_nonzero2} $\Psi=\Psi(E_-,0,\lambda)+a\{2|E|/(|E|+m)\}^{1/2}\Psi(E_-,-k,\lambda)$:
\begin{equation}
  \begin{split}
  j_x&=j_y=0,\\
  j_z&=\frac{ac^2\hbar k}{|E|+mc^2}\biggr[ a+\cos(k z-\phi) \biggr].
  \end{split}
\end{equation}
Similarly to case \ref{zero_nonzero1}, $j_z$ can be negative around $z=2n\pi/k,~n=0,\pm1,\pm2,\cdots,$ for instance at $\phi=\pi,~\lambda=1,~a=1/\sqrt{2},~m=1$.

\end{enumerate}

Some remarks are in order. First, it is notable that in case \ref{eigen} backflow always occurs with the negative-energy state. This makes sense because the negative-energy state indicates an antiparticle, which, when traveling forward, can be understood as a particle in the positive-energy state traveling backward. It also makes sense to case \ref{nonzero00}, in which the state can effectively be seen as a superposition of two components that have opposite momenta. Then, case \ref{posi-nega} is remarkable, since each component has a positive momentum (effectively), and backflow still occurs. Cases \ref{nonzero-jx} and \ref{nonzero-jx2} show that different helicities in superposition may result in nonzero current, positive or negative, in the $x$ or $y$ direction, despite that the corresponding momenta both equal zero~\cite{helicity}. Cases \ref{zero_nonzero1} and \ref{zero_nonzero2} resemble ones in the nonrelativistic limit (c.f. Eqs. (\ref{superposedstate}) and (\ref{berrysresult})) but, here, negative energies are also considered (see also \cite{melloy1998the,melloy1998probability}).

\section{Field-theoretic formalism }

In quantum field theory the wave function $\Psi$ is promoted as the field operator of fermions, which can be further expanded in terms of annihilation and creation operators satisfying fermionic anticommutation relations~\cite{weinberg}:
\begin{equation}
  \begin{split}
    &\Psi(x)=\\
    &(2\pi)^{-3/2}\sum_{\sigma}\int{\rm d\textbf{p}} \biggr[ a_{{\rm \textbf{p}},\sigma} u({\rm \textbf{p}},\sigma)e^{i p\cdot x}
    +b_{{\rm \textbf{p}},\sigma}^\dagger v({\rm \textbf{p}},\sigma)e^{-i p\cdot x}\biggr],
  \end{split}
\end{equation}
where $\{ a_{\rm \textbf{p}},a_{\rm \textbf{p}'}^\dagger \}=\{ b_{\rm \textbf{p}},b_{\rm \textbf{p}'}^\dagger \}=\delta_{\rm \textbf{p} \textbf{p}'},~\{ a_{\rm \textbf{p}},a_{\rm \textbf{p}'} \}=\{ b_{\rm \textbf{p}},b_{\rm \textbf{p}'} \}=[ a_{\rm \textbf{p}},b_{\rm \textbf{p}'} ]=[ a_{\rm \textbf{p}},b_{\rm \textbf{p}'}^\dagger ]=0$, and $u({\rm \textbf{p}},\sigma)$ and $v({\rm \textbf{p}},\sigma)$ are determined by the representation of the Lorentz group that describe the fermions with a certain spin $\sigma$ (which should not be confused with the Pauli matrices in previous sections).

Let us consider a model of spin-1/2 with the field operator distributed only around $\rm \textbf{p}=0$. We fix the spin to be upward ($\sigma=1/2$), many labels being omitted without confusion. The field operator at $t=0$ becomes then
\begin{equation}
  \begin{split}
    &\Psi(x)\biggr|_{t=0}\rightarrow(2\pi)^{-3/2}\int_{|{\rm \textbf{p}}|\leq\epsilon}{\rm d \textbf{p}}\biggr[ a_{{\rm \textbf{p}},\frac{1}{2}} u({\rm \textbf{p}},\frac{1}{2})e^{i \rm \textbf{p}\cdot \textbf{x}}\\
    &~~~~~~~~~~~~~~~~~~~~~~~~~~~~~~~~~~~+b_{{\rm \textbf{p}},\frac{1}{2}}^\dagger v({\rm \textbf{p}},\frac{1}{2})e^{-i \rm \textbf{p}\cdot \textbf{x}} \biggr]\\
    &~~\simeq(2\pi)^{-3/2}\int_{|{\rm \textbf{p}}|\leq\epsilon}{\rm d \textbf{p}}\biggr[ a_{0,\frac{1}{2}} u(0,\frac{1}{2})(1+i {\rm \textbf{p}\cdot \textbf{x}})\\
    &~~~~~~~~~~~~~~~~~~~~~~~~~~~ +b_{0,\frac{1}{2}}^\dagger v(0,\frac{1}{2})(1-i \rm \textbf{p}\cdot \textbf{x})\biggr]\\
    &~~=(2\pi)^{-3/2}\int_{|{\rm \textbf{p}}|\leq\epsilon}{\rm d \textbf{p}}\biggr[ a_{0,\frac{1}{2}} u(0,\frac{1}{2}) +b_{0,\frac{1}{2}}^\dagger v(0,\frac{1}{2})\biggr]\\
    &~~\simeq(2\pi M)^{-3/2}\int_{|{\rm \textbf{p}}|\leq\epsilon}{\rm d \textbf{p}}\biggr[a_{{\rm \textbf{p}},\frac{1}{2}}u+b_{{\rm \textbf{p}},\frac{1}{2}}^\dagger v\biggr],\label{approx}
  \end{split}
\end{equation}
where $M$, possibly depending on the small quantity $\epsilon$, equals a finite, positive constant due to the way of how the field operator is expanded around ${\rm \textbf{p}=0}$, $u$ and $v$ are short for $u(0,\frac{1}{2})$ and $v(0,\frac{1}{2})$, respectively, and
\begin{equation}
  \begin{split}
    u(0,\frac{1}{2})&=\frac{1}{\sqrt{2}}\left[ \begin{matrix}
      1\\0\\1\\0
    \end{matrix} \right],~~~~~~~~u(0,-\frac{1}{2})=\frac{1}{\sqrt{2}}\left[ \begin{matrix}
      0\\1\\0\\1
    \end{matrix} \right],\\
    v(0,\frac{1}{2})&=\frac{1}{\sqrt{2}}\left[ \begin{matrix}
      0\\1\\0\\-1
    \end{matrix} \right],~~~~~~v(0,-\frac{1}{2})=\frac{1}{\sqrt{2}}\left[ \begin{matrix}
      -1\\0\\1\\0
    \end{matrix} \right].
  \end{split}
\end{equation}
The \emph{current operator}, in analogy with (\ref{rel-current}), is then computed in the $z$-axis
\begin{equation}
\begin{split}
  &j_z\biggr|_{t=0}=(2\pi M)^{-3/2}\int_{|{\rm \textbf{p}}|\leq\epsilon}{\rm d \textbf{p}} \biggr[u^\dagger a^\dagger_{{\rm \textbf{p}},\frac{1}{2}}+v^\dagger b_{{\rm \textbf{p}},\frac{1}{2}} \biggr]\\
  &~~~~\times\left[\begin{matrix}
     &  \sigma_z\\ \sigma_z &
  \end{matrix}\right](2\pi M)^{-3/2}\int_{|{\rm \textbf{p}}|\leq\epsilon}{\rm d \textbf{p}'} \biggr[a_{{\rm \textbf{p}'},\frac{1}{2}}u+b_{{\rm \textbf{p}'},\frac{1}{2}}^\dagger v\biggr]\\
  &~~=(2\pi M)^{-3}\int_{|{\rm \textbf{p}}|\leq\epsilon}{\rm d \textbf{p}} \biggr[a_{{\rm \textbf{p}},\frac{1}{2}}^\dagger a_{{\rm \textbf{p}},\frac{1}{2}}+b_{{\rm \textbf{p}},\frac{1}{2}}b_{{\rm \textbf{p}},\frac{1}{2}}^\dagger \biggr]\\
  &~~=(2\pi M)^{-3}\int_{|{\rm \textbf{p}}|\leq\epsilon}{\rm d \textbf{p}} \biggr[a_{{\rm \textbf{p}},\frac{1}{2}}^\dagger a_{{\rm \textbf{p}},\frac{1}{2}}-b_{{\rm \textbf{p}},\frac{1}{2}}^\dagger b_{{\rm \textbf{p}},\frac{1}{2}} \biggr]\\
    &~~~~~+(2\pi M)^{-3}\int_{|{\rm \textbf{p}}|\leq\epsilon}{\rm d \textbf{p}}\cdot 1.
\end{split}
\end{equation}
Here, the first term after integration is proportional to the normal ordered charge operator $N$; the second term, which is positive and finite, depends on the way we make approximations in (\ref{approx}), but might become infinite if we perform a more general calculation with less approximations --- e.g., when $\epsilon$ is large. This is undesired, and so the physical current operator should subtract the second term~\cite{furry}:
\begin{equation}
  \begin{split}
  {\rm \textbf{j}}_z\biggr|_{t=0}&\equiv \biggr[j_z-(2\pi M)^{-3}\int_{|{\rm \textbf{p}}|\leq\epsilon}{\rm d \textbf{p}}\cdot 1\biggr]_{t=0}\\
  &=(2\pi M)^{-3}N.
  \end{split}
\end{equation}
This now has both negative and positive eigenvalues, even around ${\rm \textbf{p}}=0$. The result differs from either relativistic or nonrelativistic quantum mechanical results. (Obviously, ${\rm \textbf{j}}_{x,y}$, as well as other cases with $\sigma=-1/2$, can be computed in similar ways.) The reason for the nonzero probability current with almost vanishing momenta is that the current (\ref{rel-current}), when expressed in terms of field operators, has nothing to do with probability flows, but is a Hermitian operator which is proportional to the normal ordered charge operator.

\section{Summary}
We have investigated the quantum backflows of a free Dirac particle in superposed states with various energies, momenta, and helicities. In nonrelativistic quantum mechanics, if backflows occur, it requires necessarily that the state must involve at least two different momenta; however, in this paper, we have shown that energies and helicities, which are featured in relativistic quantum mechanics, also play an indispensable role in the occurrence of backflows. Some conditions, under which backflows occur, have no counterpart in the nonrelativistic quantum mechanics.

In the end, we have used the quantized wavefunctions (i.e., the field operators) to study the odd behaviors of the probability current at $t=0$ around ${\rm \textbf{p}}=0$. We have shown that the current operator is proportional to the normal ordered charge operator, and so it has nothing to do with any sort of probabilities. In this paper, we do not try to give an ontological interpretation of our results (except for what has been done by using the pilot-wave theory in the nonrelativistic case, Eq.~(\ref{qpotential})), because, to the best of our knowledge, most of ontological interpretations are in need of considering some presumed configuration spaces, which often suffer the localization problem when it comes to the context of relativity~\cite{localization}. Nevertheless, dynamical behaviors, as well as considering more kinds of superposed states and more general systems, may deserve subsequent investigations in the future. We suggest that the results in this paper may shed new light into relativistic electron vortices~\cite{vortex}.

\begin{acknowledgments}
    The authors wish to thank W.-Y. Hwang for helpful discussions. The project is funded by China Postdoctoral Science Foundation (Grant No. 2018M630063). J.L.C. is supported by National Natural Science Foundations of China (Grant No.\ 11475089). H.Y.S. also acknowledges the support by the Visiting Scholar Program of Chern Institute of Mathematics, Nankai University.
\end{acknowledgments}


\begin{thebibliography}{99}
\bibitem{samesign}
For the nonrelativistic particle with $E=p^2/2m$ we have ${\rm v}_p=k/2m={\rm v}_g/2$; for the relativistic particle with $E=\sqrt{p^2+m^2}$ we have ${\rm v}_p=\sqrt{k^2+m^2}/k=1/{\rm v}_g$.
\bibitem{samemass}
In quantum mechanics $\omega$ and $k$ can essentially be independent when the mass is unfixed. But we are talking about superposition here (and below) and a superposition of different masses may lead to an observable relative phase which to the best of our knowledge has not been observed so far. Thus throughout the paper we always assume tacitly that the mass should be the same in every component.
\bibitem{superpose}
The superposition of, e.g., negative and positive energies, is possible, because of a possible transition between the electron and the positron as considered in Dirac's relativistic quantum theory of the electron; see also \cite{dirac28}.
\bibitem{unorthogonal}
In case that components in superposition are not orthogonal with one another, we need to have the definition (\ref{rel-current}) divided by an addition positive factor to make the wavefunction normalized to unit. But we simply ignore such a factor (if there is indeed) because we are only interested in the directions of the current and the factor has no effect of changing them.
\bibitem{helicity}
Such a superposition of helicities is not physically impossible, because for a massive fermion its helicities are not conserved and can be transformed into one another by certain Lorentz transformations.
\end{thebibliography}
\end{document}